# Epidemic models with geography: a new perspective on r-numbers

Alan Wilson

The Alan Turing Institute

## Abstract

Most epidemic models are spatially aggregate and the index which is most used for planning and policy numbers – the r-number – typically refers to a single system of interest. Even if r-numbers are calculated for each of adjacent areas – regions or countries for example - there is no interaction between them. Here we aim to offer a fine-grained geography: models of epidemics in spatially disaggregated systems with interactions. This offers the possibility of new insights into the dynamics of epidemics and of policies aimed at mitigation and control.

## 1. Introduction.

We build on one of the standard epidemic models – SIR – which allows for 'removals' as well as infection- see Kermack and McKendrick (1927) – and we show how it can be spatially refined – characterised as 'adding geography'.

We proceed as follows. The basic model and the definition of the r-number is presented in section 2. The system with fine-grained geography added is defined in section 3, including the added complications of travel to work – and infection at residential or work location, or on public transport. The model itself is then developed in section 4 followed by concluding comments in section 5.

## 2. The basic SIR model.

Let $S(t)$, $I(t)$ and $R(t)$ be the population of susceptibles, infectives and removeds at time t, with an assumed fixed population, N, so that

$$S(t) + I(t) + R(t) = N \tag{1}$$

The dynamics of the system are described by

$$\Delta S(t, t+1) = -\lambda S(t)I(t) \tag{2}$$

$$\Delta I(t, t+1) = \lambda S(t)I(t) - \gamma I(t) \tag{3}$$

$$\Delta R(t, t+1) = \gamma I(t) \tag{4}$$

To see how the dynamics work, taking the number of infectives as an example, substitute for $S(t)$ by using (1) and then we see:

$$\Delta I(t, t+1) = \lambda\{[N - R(t) - I(t)]I(t) - \gamma I(t) \tag{5}$$

which is a form of Lotka-Volterra equation. Indeed, they could be seen as prey-predator equations, with the infected population as the predators and the susceptibles as the prey.

See Lotka (1924, 1956) for a detailed presentation, and Wilson (2006) for the principles of adding geography to these models.

We now drop the (t, t+1) and the (t) for simplicity.

Note that

$$r = \lambda/\gamma \tag{6}$$

is the much-quoted r-number – the basic reproduction number – which can be interpreted as the number of people an infective could further infect in a time period. It can be shown that if it is less than or equal to 1, I(t) declines to zero; if greater than 1, there is an epidemic. The aggregate model presented here can be highly developed to represent real situations and this will apply to the 'model with geography' that follows – but we retain the simple framework for clarity and assume that appropriate detail can be added for different circumstances.

**3. The SIR model with geography: defining the system.**

Assume a system of spatial zones labelled i = 1, 2, 3, …. - a set {i}. These could be countries, regions, cities or fine-grain zones within a city or a city region. Then our variables become, in the first instance $S_i$, $I_i$, $R_i$ with a total population of $N_i$ in zone i. We could write down versions of equations (1) – (6) simply by adding the subscript i to each variable. If sufficient and adequate data were available, the variation in $\lambda_i$, $\gamma_i$ and $r_i$ would be interesting. For example, $\lambda_i$ might be density dependent and be much lower for rural areas than for urban areas. However, to make the model more realistic by adding interaction between zones, let us assume that the population of each zone is divided into those who wholly live in i and those who live in i but also work in j (and the set {j} can include i). Assume further that the full-time residents can only be infected in i and the workers can be infected in their work location or while travelling. We introduce notation that handles this: define the variables $S_i^R$, $I_i^R$, $R_i^R$, $S_i^{RW}$, $I_i^{RW}$, $R_i^{RW}$, $S_i^{TRW}$, $I_i^{TRW}$ and $R_i^{TRW}$ for populations who are either wholly resident (R), those who are partially resident and otherwise at a work location (RW), and we identify separately, the commuting element of the workers with the additional T superscript, which will allow the probability of infection while travelling to be dependent on the length of the trip. To formulate the model, we need the number of workers travelling from each zone i to each zone j. To further simplify, assume that we know the total number of jobs at a location, $E_j^W$ that existed in the pre-infection state and that this still serves as a measure of attraction for the susceptibles who are the ones who can still travel – assuming that infectives are not fit to travel. All of these assumptions can, of course, be refined. Then, with the usual kind of spatial interaction model (Wilson, 1967, 2008), the journey-to-work flows might be:

$$T_{ij}^{RW} = A_i^{RW} S_i^{RW} E_j^W \exp(-\beta c_{ij}) \tag{7}$$

with

$$A_i^{RW} = 1/\sum_j E_j^W \exp(-\beta c_{ij}) \tag{8}$$

The total number of work-based susceptibles at j is defined as $S_j^W$:

$S_j^W = \sum_i T_{IJ}^{RW}$ (9)

This produces a measure of suscepibles who work in a zone which is a function of all zones through those who commute into the zone – through equations (7) and (8). For convenience, we can take the mirror image of equation (9) so that we can collect together the different variables for zone i:

$S_i^W = \sum_j T_{ji}^{RW}$ (9')

## 4. The disaggregated model.

We can then identify the changes in the numbers of susceptibles, infectives and removeds in each zone i. It is essential that we specify carefully where the infection takes place in the time period, particularly distinguishing workplace and travel to avoid double counting. Full-time residents are infected in the zone and the workers in a combination of travel and workplace zones. For simplicity, we assume that for workers, infection during travel precedes infection at the workplace and those infected are removed from the number of susceptibles in the workplace. This is very much an approximation of course and can be refined as appropriate. Residents of zone 1 can be infected in three kinds of location: in zone i itself for full-time residents, or while travelling to work, or at workplaces. Residents, commuters and workers are assumed to have different rates of infection. The dynamic equations become [adding space to equations (2) – (4)]:

$\Delta S_i^R = -\lambda_i^R S_i^R I_i^R$ (10)

$\Delta I_i^R = \lambda_i^R S_i^R I_i^R - \gamma I_i^R$ (11)

$\Delta R_i^R = \gamma I_i^R$ (12)

Because we are assuming that commuters are infected ahead of workplace infection (notwithstanding the complexities of two-way trips that can be incorporated as a later refinement), we take the TRW category next:

$\Delta S_i^{TRW} = -\lambda_i^{TRW} \sum_j T_{ij}^{RW} \cdot I_{ij}^{RW}$ (13)

$\Delta I_i^{TRW} = \lambda \sum_j T_{ij}^{RW} \cdot I_{ij}^{RW}$ (14)

$\Delta R_i^{TRW} = \gamma \sum_j I_{ij}^{TRW}$ (15)

We can then deal with the RW category, reducing the susceptibles to allow for travel infection (and noting that that is the only variable of the right hand side that is (t, t+1) rather than t):

$\Delta S_i^{RW} = -\lambda_i^{RW} [\sum_j T_{ji}^{RW} - \Delta S_i^{TRW}]_i^{RW}$ (16)

where here and in the following equation, $\Delta S_i^{TRW}$ is taken from equation (13)

$\Delta I_i^{RW} = \lambda_i^{RW} [\sum_j T_{ji}^{RW} - \Delta S_i^{TRW}]_i^{RW} - \gamma I_i^{RW}$ (17)

$\Delta R_i^{RW} = \gamma I_i^{RW}$ (18)

Simulation tests with this model should show interesting dynamics. It would be possible, for example to start at t=0 and locate some initial infections, and then simulate the spread, taking account for example of the λs being density dependent. It should be possible to interpret the individual r-numbers. It should be possible to identify critical points (r-numbers or λs) at which epidemics break out locally.

A first step will be to estimate an r- number for a zone – first constructing each $\lambda_i$ as an average:

$$\lambda_i = (\lambda_i^R I_i^R + \lambda_i^{RW} I_i^{RW} + \lambda_i^{TRW} I_i^{TRW}) / (I_i^R + I_i^{RW} + I_i^{TRW}) \tag{19}$$

and then, as usual,

$$r_i = \lambda_i / \gamma \tag{20}$$

Any one of these zonal r-numbers are dependent on the other zones through the spatial interaction represented by the commuting equation (7). Of course, this aspect of the formulation could be extended to include other aspects of spatial interaction – from migration at a coarse scale to different kinds of trip purposes at a finer scale.

A system-wide λ might be

$$\lambda = \sum_i (\lambda_i^R I_i^R + \lambda_i^{RW} I_i^{RW} + \lambda_i^{TRW} I_i^{TRW}) / \sum_i (I_i^R + I_i^{RW} + I_i^{TRW}) \tag{21}$$

A conjecture to be tested is then: is λ = 1 still a critical point? We can interpret and explore the $\{\lambda_i\}$ by zone; and the disaggregated λs - $\lambda_i^R$, $\lambda_i^{RW}$, $\lambda_i^{TRW}$. We can check for internal consistency – for example against double counting - by making all the λ parameters equal and explore the result. This would not collapse to the aggregate system because the spatial interaction model [equations (7) and (8)] removes any symmetries.

**5. Further disaggregation: towards realism.**

Infection rates will differ in different kinds of workplaces. Suppose we want to identify schools, hospitals and care homes for example; we might also want to distinguish office, retail and construction workers before we add a residual 'other' class for completeness. We can develop our notation to add this layer of disaggregation. In the equations in section 4 above, W is used as a superscript to identify a work population or flow. Now let w be a category within the working population and then, formally,

$$E_j^W = \sum_w E_j^w \tag{22}$$

To avoid further complications in notation at this stage, we should treat school pupils as 'workers' and be one of the w-classes. Adult school workers could be a separate group or simply included in this category. We then need to ensure that bour core assumption is in place: that in the time period, a worker can only be infected in one place. With this formulation, we can introduce the following modifications to the model in section 4.

(a) We can retain equations (13) – (15) as an approximation; or alternatively, replace W by w throughout and have a journey to work model for each w-category. At some point, ideally, we should also distinguish transport mode.

(b) In equations (16) – (18), we replace W by w, thus having a set of w-equations with $\lambda_i^{Rw-}$ as a set of infection rates for each w class. This would allow us to have higher rates for workers in, and visitors to, care homes for example.

(c) It might also be worthwhile to modify equation (10) by replacing $I_i^R$ by $(I_i^R + {}_i^{RW})$ to allow for workers returning home from high infection institutions and areas to add to residential infection.

Of course, there could me many other avenues to disaggregation – notably age and, as already noted, transport mode. As was argued in traditional maths text books, this is left for the time being as an exercise for the reader!

The level of disaggregation suggested here would allow for higher ongoing transmission from hotspots in simulating the spread of an academic with implications for mitigation nand post-lockdown policies.

### 6. Concluding comments.

This a simple model to illustrate an important point: that infection rates will vary geographically and by person type – here, worker or non-worker - but could be extended in various ways. There is little doubt that the usual r-value calculated for an aggregate system is made up of a number of elements in a system and the exploration of the more complex representation, incorporating person-type, space and different kinds of activities, could inform, for example, policies on the treatment of hotspots or the relaxation of lockdowns. Our ability to do this will depend on the availability of appropriate data, or the deployment of methods such as microsimulation or bi-proportional fitting to generate 'data' from samples.

May 2020